\documentclass[epj,nopacs]{svjour}

\title{The gravity/CFT correspondence}
\date{}
\author{ Henrique Gomes\inst{1} \and Sean Gryb\inst{2} \and Tim Koslowski \inst{3} \inst{4} \and Flavio Mercati\inst{3} \inst{5} \inst{6}}

\institute{University of California at Davis, One Shields Avenue Davis, CA, 95616, USA \and 
    Utrecht University,  Leuvenlaan 4, 3584 CE Utrecht \and
    Perimeter Institute for Theoretical Physics, Waterloo, Ontario N2L 2Y5, Canada \and
    Dept. of Mathematics and Statistics, University of New Brunswick, Fredericton, New Brunswick, Canada, E3B 5A3 \and
    Departamento de F\'isica Te\'orica, Universidad de Zaragoza, Zaragoza 50009, Spain \and
    School of Mathematical Sciences, University of Nottingham, Nottingham NG7 2RD, United Kingdom
}

\abstract{General Relativity can be formulated in terms of a spatially Weyl invariant gauge theory called Shape Dynamics. Using this formulation, we establish a ``bulk/bulk'' duality between gravity and a Weyl invariant theory on spacelike Cauchy hypersurfaces. This duality has two immediate consequences: i) it leads trivially to a corresponding ``bulk/boundary'' duality between General Relativity and a boundary CFT, and ii) the boundary can be defined in a gauge--invariant way. Moreover, the corresponding bulk/boundary duality is sufficient to explain a large portion of the evidence in favor of gauge/gravity duality and provides independent evidence for the AdS/CFT correspondence.}

\usepackage{amssymb}
\usepackage{amsmath}
\usepackage{graphicx}
\usepackage[normalem]{ulem}

\def\lf {\ensuremath{\left}}
\def\rt {\ensuremath{\right}}

\def\hg {\ensuremath{\mathcal H_\text{gl}}}

\let\oldmarginpar\marginpar
\renewcommand\marginpar[1]{\oldmarginpar{\color{red}\raggedright\scriptsize #1}}

\newcommand{\eq}[1]{(\ref{eq:#1})}

\newcommand{\order}[1]{\ensuremath{\mathcal{O}(#1)}}
\newcommand{\diby}[2]{\ensuremath{\frac{\delta #1}{\delta #2}}}

\newcommand{\pb}[2]{\ensuremath{\lf\{#1,#2 \rt\}}}

\newcommand{\mean}[1]{\ensuremath{\lf\langle #1 \rt\rangle }}

\newcommand{\hn}[1]{\ensuremath{\mathcal H_{(#1)}}}
\newcommand{\wn}[1]{\ensuremath{\omega_{(#1)}}}
\newcommand{\sn}[1]{\ensuremath{S_{(#1)}}}

\begin{document}

\maketitle


\section{Introduction}

The AdS/CFT correspondence conjectures an identification between certain quantities in classical asymtotically anti--de Sitter (AdS) gravity and quantities in a conformal quantum field theory (CFT) at the asymptotic boundary\cite{Maldacena,Witten,LargeReview}. This bulk/boundary correspondence is usually understood by observing that the asymptotic symmetry group of asymptotically AdS gravity can be identified with the symmetry group of a CFT. The original AdS/CFT conjecture related particular asymptotics of type $II\,B$ string theory and $N=4$ super Yang--Mills theory, but there is a large body of literature that provides an unprecedented amount of evidence for a much more general gauge/gravity correspondence (see \cite{Horowitz:gauge_gravity_review} for a discussion) through explicit calculations in which the radial evolution of the classical gravity theory near the AdS boundary is compared with the RG flow of a conformal field theory at the boundary. This general correspondence has found applications in condensed matter systems, quark--gluon plasmas, and black hole physics \cite{Wadia:gauge_gravity_appliations}.

The large number of papers makes it virtually impossible to give a fair account of the current status of this vast field of research, although a classic review can be found in \cite{LargeReview} and a modern introduction can be found in \cite{Polchinski:intro_gauge_gravity}. It is, however, fair to say that there are two points generically assumed in the AdS/CFT literature: the correspondence
\begin{enumerate}
    \item is bulk/boundary, and
    \item relates gravity observables to CFT operators.
\end{enumerate}
By \emph{bulk/boundary} we mean that the near boundary, or large radius, behavior of gravity is dual to a CFT as the radius is taken to infinity.

This brings us to the point of this paper: using a novel formulation of General Relativity (GR), called Shape Dynamics (SD), we
\begin{enumerate}
    \item prove a \emph{bulk/bulk} duality, and
    \item give a one--to--one relation between \emph {all} observables of classical gravity and a classical Weyl invariant field theory.
\end{enumerate}
By \emph{bulk/bulk} duality we mean that the behavior of gravity on \emph{any} Cauchy surface in the bulk -- satisfying the Constant Mean Curvature (CMC) condition -- is dual to a Weyl invariant theory on that surface. We then show how the bulk/bulk duality between GR and SD straightforwardly reduces to a bulk/boundary duality reminiscent of AdS/CFT.

This is achieved by considering a natural expansion parameter for near--boundary solutions of SD: the spatial volume of the Universe. We recover the same expansion as in previous computations of AdS/CFT and holographic renormalization \cite{Verlinde,Skenderis:holoRG_main,Skenderis:holo_RG,Skenderis:holo_weyl} but our framework has two important advantages: i) the appearance of Weyl invariance at the boundary is a trivial consequence of bulk/bulk duality, and ii) the expansion parameter in our theory is gauge invariant. This is not true of the expansion parameter used in standard holographic renormalization. These advantages are a direct result of SD having a different symmetry (i.e., spatial Weyl invariance) than GR (i.e., foliation invariance), while still describing the same physics.

Using a standard semiclassical treatment, we provide a relation between classical Shape Dynamics in the bulk and a 3 dimensional semiclassical QFT near a fixed point. We then make a key observation: \emph{these results could serve as an explanation for a large part of the theoretical evidence in favor the AdS/CFT conjecture.} Such an independent explanation could be very important should it turn out that the assumptions underlying the precise AdS/CFT correspondence can not be fulfilled.

The motivation for SD comes from Machian principles, which are rooted in the observation that all length measurements are local comparisons. This suggests that local spatial scale should be pure gauge, which is neither realized in GR nor in any of its gauge fixings. However, SD exploits a particular gauge--\emph{un}fixing of GR in CMC gauge to implement Weyl invariance. Thus, SD is, by construction, dynamically equivalent to GR, but possesses local spatial Weyl invariance as a gauge symmetry.  By replacing local refoliation invariance in GR with local Weyl invariance, SD solves the problem of many fingered time in canonical gravity. However, we will see that equivalence of the dynamics of GR and SD requires a non--local Hamiltonian. It was first found in 2+1 dimensions that this Hamiltonian becomes a local phase space function in a regime where the CMC volume dominates over all other degrees of freedom \cite{Budd:2_plus_1_sd}. We will show that performing a large volume expansion of the Shape Dynamics Hamiltonian resembles many features of the familiar near boundary expansion, despite technical differences in the construction. It is also worthwhile to note that since our aim here is to observe that the duality exists irrespective of the AdS/CFT correspondence, we have restricted ourselves to 3+1 spacetimes with positive cosmological constant. The generalization to different dimensions and values of the cosmological constant is straightforward.

We start with a brief description of SD to explain why the classical correspondence is bulk/bulk. We proceed with the large volume expansion of the SD Hamiltonian to explicitly show that the Hamiltonian is local\footnote{By ``local'' we mean involving only a finite number of derivatives.} in this regime. Subsequently we consider an approximation to Hamilton's principal function, which we use to discuss the relation of a semiclassical SD wavefunction and the $\hbar$ expansion of a hypothesized CFT partition function.

\section{Shape Dynamics}

SD is a Hamiltonian theory on the phase space of Arnowitt--Deser--Misner (ADM)  \cite{ADM}. This phase space consists of a 3--dimensional metric $g_{ab}$ and its conjugate momentum density $\pi^{ab}$. The generalization to higher dimensions is straightforward. For simplicity, we assume a compact Cauchy surface without boundary (asymptotically flat SD has been worked out in \cite{Gomes:linking_paper}). To show the equivalence, we will first describe a rigorous procedure that links the two theories then describe the intuitive picture that emerges. For a more thorough account see \cite{Gomes:sd_paper,Gomes:linking_paper} or \cite{gryb:phd,gomes:phd} for a more pedagogical exposition.

The fastest way to construct SD is to consider a \emph {linking theory}, which is obtained by adjoining a scalar field, $\phi$, and its canonically conjugate momentum density, $\pi_\phi$, to the ADM phase space. We then adjoin the constraint 
\begin{equation}
 Q(\rho)=\int d^3x \, \rho \, \pi_\phi,
\end{equation}
which states that $\phi$ is pure gauge, to the ADM constraints
\begin{eqnarray}
   H(\xi)&=&\int d^3x\, \pi^{ab}\mathcal L_\xi g_{ab} \\
   S(N)&=&\int d^3x\, N\left(\frac{1}{\sqrt{g}}\pi^{ab}G_{abcd}\pi^{cd}-\left(R(g)-2\Lambda\right)\sqrt{g}\right).\nonumber
\end{eqnarray}
The field $\phi$ behaves like a conformal compensator parametrizing an artificial conformal invariance. $\xi^a(x)$, $N(x)$ and $\rho(x)$ denote Lagrange multipliers, $\mathcal L_\xi$ is the Lie derivative along a vector field $\xi$, $G_{abcd} = g_{ac} g_{bd} - \frac 1 2 g_{ab} g_{cd}$ is the inverse of the DeWitt supermetric, $R$ is the Ricci scalar, and $\Lambda$ the cosmological constant. The conformal compensator is introduced through a canonical transformation, $T$, that performs conformal transformations, parametrized by $\phi$, on $g_{ab}$ that preserve the total spatial volume. $T$ is defined as
\begin{equation}\label{equ:canonicalTrf}
 \begin{array}{rcl}
   g_{ab}&\to&e^{4\hat \phi}g_{ab}\\
   \pi^{ab}&\to&e^{-4\hat \phi}\left(\pi^{ab}- \frac 1 3\langle\pi\rangle\left(1-e^{6\hat \phi}\right)g^{ab}\sqrt{g}\right)\\
   \phi&\to&\phi\\
   \pi_\phi&\to&\pi_\phi-4\left(\pi-\langle\pi\rangle\sqrt{g}\right),
 \end{array}
\end{equation}
where $\pi=\pi^{ab}g_{ab}$ and the mean operator is defined by $\langle f\rangle\equiv\frac 1 V\int d ^3 x\sqrt{g} f$ for arbitrary functions $f$ and without $\sqrt{g}$ for densities. The `hat' operator enforces that the conformal transformations are volume preserving. Thus, it must be such that
\begin{align}\label{equ:phiNormalization}
    V[g] &= V[ Tg ], \text{ or} & \left\langle e^{6\hat\phi}\right\rangle&=1.
\end{align}
It can be explicitly parametrized by $\hat \phi=\phi-\frac 1 6 \ln\langle e^{6\phi}\rangle$. 

To see that the canonically transformed theory is equivalent to GR, we impose the gauge fixing condition $\phi \equiv 0$ and perform phase space reduction. This leads immediately to GR by returning the ADM constraints to their original form and by fixing the Lagrange multiplier $\rho\equiv 0$. SD is obtained by imposing the gauge fixing condition $\pi_\phi\equiv 0$. This turns the constraints $Q$ into the generators
\begin{equation}
 D=\pi-\langle\pi\rangle\sqrt{g}
\end{equation}
of spatial conformal transformations that preserve the total spatial volume. It can be shown, \cite{Gomes:sd_paper,Gomes:linking_paper}, that $\rho\equiv 0$ is a partial gauge fixing of $S(N)$ that fixes all but the freedom to redefine a global normalization of $N$. Because of the $D$ constraints, this corresponds to a choice of lapse where the hypersurfaces are CMC.
The phase space reduction is completed by treating $TS = 0$ as an equation for $\phi$. This means we must simultaneously solve
\begin{equation}\label{equ:SD-definition}
 \begin{array}{rcl}
   \mathcal \hg [g,\pi]&=&\frac 1 {\sqrt{g}} TS[g,\pi;x)\\
   1&=&\left\langle e^{6\phi[g,\pi,x)}\right\rangle
 \end{array}
\end{equation}
for the field $\phi(x)= \phi_o[g,\pi,x)$ and the spatial constant $\hg$. A positive solution for $\phi$ is known to exist and to be unique. The result is the total SD Hamiltonian
\begin{equation}
 H_{SD}=H(\xi)+D(\rho)+\mathcal N \hg,
\end{equation}
where the vector field $\xi$ is the Lagrange multiplier of the usual diffeomorphism constraints, $\rho$ the Lagrange multiplier of the volume preserving conformal constraints, and $\mathcal N$ the Lagrange multiplier of $\hg$.

\begin{figure}[h!]
\begin{center}
\includegraphics[width=0.45\textwidth]{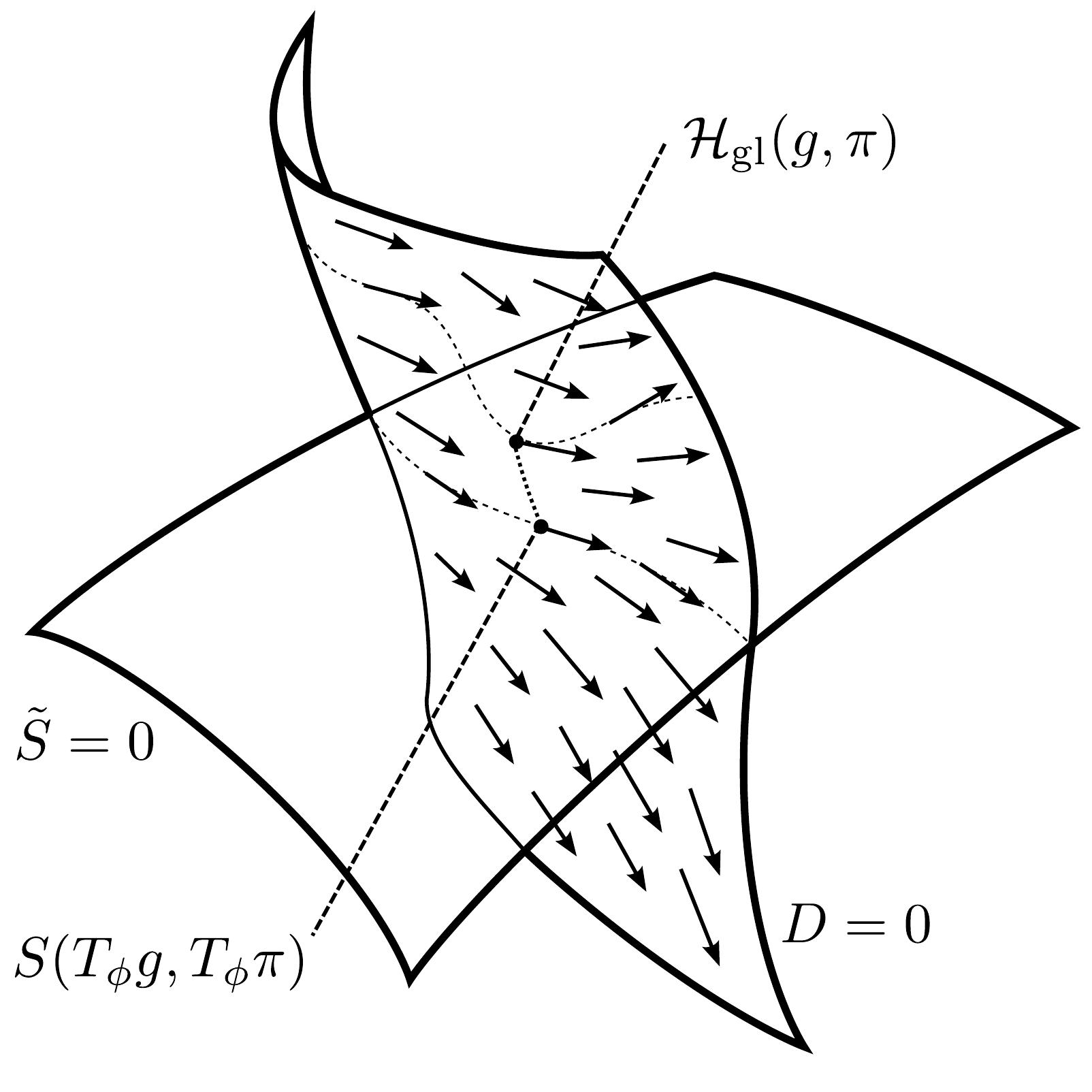}
\caption{The definition of $\hg$. The constraint surface $D\approx 0$ intersects $\tilde S \approx 0$. The Hamilton vector field of $\hg$ is defined by the value at the gauge fixed surface, represented by the dark dotted line.}\label{fig:sd}
\label{default}
\end{center}
\end{figure}

We can now describe a picture, see Fig (\ref{fig:sd}),
that provides an intuitive shortcut in relating GR to SD. Both theories are defined by constraint surfaces on the ADM phase space. Because $D=0$ is a proper \emph{partial} gauge fixing of $S$, there is a subset, $\tilde S$, of $S$ whose surface $\tilde S = 0$ intersects exactly once with the constraint surface, $D=0$, of SD (recall that the constraint surface of the diffeomorphism constraint is shared by both theories). Along this intersection, $\hg$ is defined as the remaining non--zero part of $S$. This definition will ensure that the trajectories of both theories are the same on this intersection. To define $\hg$ everywhere on $D=0$, we must lift its Hamilton vector flow along the gauge orbits of $D$. Since these are the volume preserving conformal transformations, this leads directly to our result that $\hg[g,\pi] = \frac 1{\sqrt g} TS[g,\pi,x)$. Note that, by definition, $\hg$ Poisson comutes with the $D$'s.

The inclusion of matter into the SD formalism is discussed in \cite{MatterPaper,Gomes:matter_coupling_proc,Gomes:h_gl_uniqueness}. The procedure is relatively straightforward and follows from the results of \cite{Isenberg:matter_coupling1,Isenberg:matter_coupling2}. Many forms of matter can be included but there are interesting subtleties discussed in detail in \cite{Gomes:h_gl_uniqueness}. In this paper, we will only consider the pure gravity sector of the correspondence, for which interesting conclusions can already be drawn. A more complete analysis that includes matter is currently underway.

To use the bulk/bulk equivalence of GR and SD to study the bulk/boundary correspondence, we proceed as follows: i) we perform a large volume expansion of the SD Hamiltonian, and ii) we consider large volume boundary conditions for the principal function that are compatible with an asymptotically homogeneous space.

\section{$V$ expansion}

It will be convenient to take advantage of the fact that $R$ can be chosen to be a spatial constant $\tilde R$ by the conformal transformation
\begin{equation}
  \hat  \phi \to  \hat \phi +  \hat \lambda,
\end{equation}
where $\lambda$ is the solution to the Yamabe problem. To expand $\hg$ in powers of $V^{-2/3}$, the explicit $V$ dependence of $\hg$ must be isolated. This can be done using the change of variables $(g_{ab};\pi^{ab}) \to (V,\bar g_{ab}; P, \bar \pi^{ab})$ given by
\begin{equation}
  \begin{array}{rl}
       \bar g_{ab} &= \left(\frac{V}{V_0} \right)^{-\frac{2}{3}} g_{ab}, \qquad \qquad V = \int d^3x \sqrt{g}, \\
       \bar \pi^{ab} &= \left(\frac{V}{V_0}\right)^{\frac{2}{3}}\left(\pi^{ab} - \frac 1 3 \mean{\pi} g^{ab} \sqrt{g} \right),  ~ P= \frac 2 3 \mean{\pi}.
  \end{array}
\end{equation}

Where $V_0 = \int d^3 x \sqrt{\bar g}$ is a fixed reference volume, which can be fixed by choosing a volume where the systems enters an expanding regime. As we will see, the large volume limit is an attractive fixed point of classical dynamics. One can easily verify that $\pb{V}{P} = 1$ and $\pb{P}{\bar g_{ab}} = \pb{P}{\bar \pi^{ab}} = 0$, so that the barred variables are independent of $V$. Note that $\bar g$ and $\bar \pi$ are not canonically conjugate.

Using these variables, calling $\Omega = e^{\hat\phi+ \hat\lambda}$, inserting (\ref{equ:canonicalTrf}) into (\ref{equ:SD-definition}), and repeatedly using $D = 0$, we can drastically simplify the expression in (\ref{equ:SD-definition}). Using these simplifications, we see that $\hg$ is found by simultaneously solving the equations
\begin{eqnarray}\label{eq:main hg}
\begin{array}{r}
   \hg = \lf( 2\Lambda - \frac{3}{8}P^2 \rt) + \frac{\lf( 8 \bar{\nabla}^2 - \tilde R \rt) \Omega}{(V/V_0)^{2/3}\Omega^5}
       - \frac{\bar \pi^{ab} \bar \pi_{ab} }{(V/V_0)^2\Omega^{12}\bar g}
     \end{array}
   \\
  \mean{\Omega^6} = 1, \label{eq:Omega}
\end{eqnarray}
where barred quantities are calculated using $\bar g_{ab}$. Perform the expansion ansatz
\begin{equation}
   \hg = \sum_{n=0}^\infty  \left(\frac{V}{V_0} \right)^{-2n/3} \hn{n}, ~~ \Omega^6 = \sum_{n=0}^\infty  \left(\frac{V}{V_0} \right)^{-2n/3}  \wn{n} .
\end{equation}
The restriction \eq{Omega} is trivially solved by $\mean{\wn{n}} = 0$ for $n \neq 0$ and $\mean{\wn{0}} = 1$. We can solve for the $\hn{n}$'s by inserting the expansion, taking the mean, and using the fact that $\tilde R$ is constant. The solution for $\hn{n}$ can be used to solve recursively for $\wn{n}$. For a detailed outline of the technical steps in this expansion, see \cite{gomes:phd,gryb:phd}. The first non--vanishing terms are
\begin{eqnarray}\label{eq:hg v exp}
   \hg &=& \lf( 2\Lambda - \frac 3 8 P^2 \rt) - \left(\frac{V_0}{V}\right)^{2/3}  \mean{\tilde R}\nonumber         \\
&&           + \lf(\frac{V_0}{V}\rt)^2 \mean{\frac{\bar \pi^{ab} \bar \pi_{ab}}{\bar g}} + \order{\lf(V/V_0\rt)^{-8/3}},
\end{eqnarray}
where the $(V/V_0)^{-4/3}$ term is explicitly found to vanish. The next order terms are significantly more complex and non--local as they involve the inverse Laplacian. The first term of (\ref{eq:hg v exp}) implies that, in the large volume limit, the Hamiltonian constraint, $\hg = 0$, takes the simple form
\begin{equation}
    P = \pm \sqrt{ \frac{16} 3 \Lambda}.
\end{equation}
Since $P$ generates translations in $V$, the expanding solution is an \emph{attractive} fixed point of the classical dynamics.

The expression \eq{hg v exp} combined with the conformal constraints allows us to make the connection with CFT. Using $P = \frac 2 3 \mean{\pi}$, solving $\hg = 0$ for $\mean{\pi}$, and adding the result to $D\approx 0$, we obtain
\begin{equation}
   \pi / \sqrt{g} = \pm \, c.
\end{equation}

\section{Hamilton--Jacobi (HJ) equation}

We can now solve the HJ equation for SD in the large volume limit. The HJ equation can be obtained from \eq{hg v exp} by making the substitutions
\begin{align}
   P &\to \diby{S}{V} & \pi^{ab} &\to \diby{S}{ g_{ab}},
\end{align}
where $S= S(g_{ab}, \alpha^{ab})$ is the HJ functional that depends on the metric $g_{ab}$ and parametrically on the separation constants $\alpha^{ab}$. These integration constants are symmetric tensor densities of weight 1. We can express $\bar \pi^{ab}$ in terms of $\diby{S}{ g_{ab}}$ and use the chain rule to write the result in terms of $\diby{S}{V}$ and $\diby{S}{\bar g_{ab}}$. The $V$ derivatives drop out of the final expression, which is
\begin{equation}
   \bar \pi^{ab} \to \diby{S}{\bar g_{ab}} - \frac 1 3 \mean{ \bar g_{ab} \diby{S}{\bar g_{ab}}} \bar g^{ab} \sqrt{\bar g}.
\end{equation}
We expand $S$ in powers of $(V/V_0)^{-2/3}$,
\begin{equation}
   S = \sum_{n=0}^\infty \lf( \frac{V}{V_0}\rt)^{(3-2n)/3} \sn{n}
\end{equation}
and we insert this expansion into the HJ equation obtained using the substitutions above. To obtain a complete integral of the HJ equation, $\sn{0}$ can be taken of the form $\sn{0}=\int d^3 x \alpha^{ab}g_{ab}$. The linear constraints determine $\alpha^{ab}$ to be transverse and with covariantly constant trace. The leading order HJ equation determines the value of the trace of $\alpha^{ab}$. This restricts the freely specifiable components of $\alpha^{ab}$ precisely to the freely specifiable momentum data in York's approach \cite{York:cotton_tensor}.

For this paper, we will restrict ourselves to separation constants with vanishing transverse traceless part.\footnote{These conditions are compatible with asymptotic (in time) dS space, which has maximally symmetric CMC slices.} The treatment of general separation constants is vastly more complicated and will not lead to the homogeneous spaces that we are interested in. We thus restrict our attention to the initial condition
\begin{equation}
   \sn{0} = \pm \sqrt{ \frac {16} 3 \Lambda }\, V_0.
\end{equation}

To obtain the remaining $\sn{n}$'s, we can use the gauge invariance of $\hg$ under the action of the $D$'s to work in a gauge where $R$ is constant. In this gauge, $\tilde R = R$. The variations of $\tilde R$ can be found in a derivative expansion using the standard variations of $R$. The first few orders of the derivative expansion of the first $\sn{n}$'s are not affected by $\lambda$. These are
\begin{align}
\label{1}   \sn{1} &= \mp \sqrt{\frac 3 \Lambda }  \tilde R \, V_0 = \mp \sqrt{\frac 3 \Lambda }  \int d^3x \sqrt {\bar g} \tilde R, \\
\label{2}   \sn{2} &= \pm \lf(\frac 3 \Lambda\rt)^{3/2} \int d^3 x \sqrt {\bar g} \lf( \frac 3 8 \tilde R^2 - \tilde R^{ab} \tilde R_{ab} \rt).
\end{align}
The higher order terms can be obtained straightforwardly but become increasingly more involved because of the non--local terms appearing in the V expansion of $\hg$. Gauge invariant solutions can be obtained by restoring the $\lambda[g,x)$ dependence of the tilded variables. This solves the local HJ constraints of SD for asymptoticly homogeneous initial conditions.

In a semiclassical approximation, the phase of the wavefunctional is given by the solution to the HJ equation, which we want to relate to the partition function of a conformal field theory: $\Psi_\text{sd} = \Psi_\text{sd}^+ + \Psi_\text{sd}^- =a_+ e^{\frac i \hbar S_+} + a_- e^{\frac i \hbar S_-}$, where $S_\pm$ represent the expanding and contracting solutions of the HJ equation. Using the gauge invariance of our solutions under volume preserving conformal transformations, it follows, by differentiating with respect to the volume, that the $\Psi_\text{sd}^\pm$ obey the identities
\begin{equation}
   i\hbar g_{ab} \frac 1 {\sqrt g} \diby{}{g_{ab}} \Psi_\text{sd}^\pm(g) = \mp A(g) \Psi_\text{sd}^\pm(g),
\end{equation}
where
\begin{eqnarray} \label{eq:conf_ano}
&&   A(g) = \sqrt{ \frac {16} 3 \Lambda } - \sqrt{\frac 1 {3\Lambda} } \mean{\tilde R} ~ \lf( V/V_0\rt)^{-2/3} \\
&&- \frac 1 3 \lf(\frac 3 \Lambda\rt)^{3/2} \mean{\frac 3 8 \tilde R^2 - \tilde R^{ab} \tilde R_{ab}} ~ \lf( V/V_0\rt)^{-4/3}+ \cdots \nonumber
\end{eqnarray}
If we now take the expanding solution $\Psi_\text{sd}^+$ and reinterpret it as the partition function of a CFT $Z=\Psi_\text{sd}^+$, we get a semiclassical expression expression that resembles a conformal Ward identity.\footnote{In even dimensions, there would be a (finite) V--independent piece that resembles a conformal anomaly.} 

The advantage of SD is that the local constraints are \emph{linear} in the momenta. This implies that the SD constraints can be quantized unambiguously as vector fields on configuration space leading to linear constraints on a semiclassical wave function. It follows that the wavefunctional of SD is invariant under diffeomorphisms and volume preserving conformal transformations. The correspondence thus implies that the CFT partition function is also invariant under diffeomorphisms and volume preserving conformal transformations at all values of the CMC volume and not just in the infinite volume limit.

This correspondence suggests an interesting possibility: the potential of developing a construction principle for SD that does not rely on having GR at our disposal. Such a construction principle might be obtained through the correspondence by trying to implement Barbour's understanding of time \cite{barbour:eot}. The identification of ``volume time'' with ``RG time'' suggests that time is identified with the level of coarse graining of a CFT. Coarse graining is roughly a restriction of the complexity of configuration space. If true, this would imply that time is given by a measure for complexity. Thus, this construction principle for the SD Hamiltonian would provide a realization of Barbour's idea that the flow of time enters a timeless theory through a measure of complexity. He calls this measure the abundance of ``time capsules.''

Our derivation is close in spirit to \cite{Verlinde} and is particularly inspired by Freidel \cite{Freidel}. As mentioned, being able to impose all local constraints is not explicitly viable in GR, and thus provides an advantage of the SD approach. Our work also sheds a different light on the status of bulk--gravity/boundary--CFT correspondences by identifying the classical bulk/bulk equivalence rather than particular boundary conditions as the source of the correspondence. This means in particular that there is a classical explanation of dualities of the type of semiclassical AdS/CFT which does not require the full AdS/CFT correspondence to hold.

Our assumptions for the construction of the large volume SD Hamiltonian are compact CMC slices and the existence of trajectories that reach the large volume limit. Thus, we have shown that the correspondence from SD does not need to assume asymptotic (A)dS space, but is a generic large CMC volume gravity/CFT correspondence.  This lies in contrast to the usual derivation of the Hamilton--Jacobi expansion (see  \eqref{1} and \eqref{2}) in the holographic AdS/CFT literature \cite{Skenderis:holoRG_main}. There, it is assumed that spacetime is asymptotically AdS, in a well--defined sense which implies that one can use the geodesic distance from the conformally compactified boundary (in the fixed bulk metric),  $r$, as an expansion parameter. In these coordinates, the near boundary metric becomes $ds^2=\frac{1}{r^2}(dr^2+g_{ij}(x,r)dx^idx^j)$ and one bases the construction on an expansion of the spatial metric $g_{ij}(x,r)$ in $r$. The corresponding  assumption on the asymptotic behavior of the extrinsic curvature allows one to identify the dilatation operator with the radial derivative. If one further assumes that the metric and extrinsic curvature are analytic in this expansion parameter, then, after expanding the extrinsic curvature in terms of dilatational eigenvalues, one can use the identification between the dilatation operator and the radial derivative and the Einstein equations to  reconstruct the on--shell action in terms of an $r$ expansion.
 
Although the Hamilton--Jacobi action functional we arrive at is the same as in \cite{Skenderis:holoRG_main}, the method and computations are vastly different. Let us pause to take stock of the different  requirements of both approaches: in the holographic renormalization procedure, one requires a fixed background structure present in the expansion parameter, which is given by the geodesic distance of an unperturbed metric. One furthermore requires that the radial derivative equals the dilation operator close to the boundary and that all quantities be analytic in $r$.

In contrast, we require that the Universe approaches a very large volume, using the volume itself as our expansion parameter. Since the volume of the Universe is invariant under our gauge transformations, it is a physically meaningful parameter under which to perform an expansion.  That both approaches yield the same result could be explained by the fact that the temporal gauge $N=1$ used in the usual holographic renormalization procedure approximates a CMC gauge, where the theory approaches Weyl invariance, for very large radii.
In light of these advantages, we believe that SD may be the natural framework for further exploring the connection between gravity in the large CMC volume limit and boundary CFT.

\begin{acknowledgement}

We would like to thank Julian Barbour for his unique vision and perseverance towards finding a scale--free description of gavity, Lee Smolin for encouraging us to explore the link between shape dynamics and the gravity/CFT correspondence, and Laurent Freidel for useful discussions. Research at the Perimeter Institute is supported in part by the Government of Canada through NSERC and by the Province of Ontario through MEDT. This work was funded, in part, by a grant from the Foundational Questions Institute (FQXi) Fund, a donor advised fund of the Silicon Valley Community Foundation on the basis of proposal FQXi-RFP2-08-05 to the Foundational Questions Institute. HG was supported in part by the U.S. Department of Energy under grant DE-FG02-91ER40674.

\end{acknowledgement}

\bibliographystyle{utphys}
\bibliography{BIB_DS_CFT.bib}

\end{document}